\documentclass[sigconf,screen]{acmart}
\AtBeginDocument{%
  }

\graphicspath{{figures/}}

\copyrightyear{2026}
\acmYear{2026}
\setcopyright{cc}
\setcctype{by}
\acmDOI{10.1145/3770743.3804116}
\acmConference[DAC '26]{63rd ACM/IEEE Design Automation Conference}{July 26--29, 2026}{Long Beach, CA, USA}
\acmBooktitle{63rd ACM/IEEE Design Automation Conference (DAC '26), July 26--29, 2026, Long Beach, CA, USA}
\acmISBN{979-8-4007-2254-7/2026/07}

\usepackage{multirow}
\usepackage{booktabs}
\usepackage{stfloats}
\usepackage{threeparttable}
\begin{document}

\title{CapBench: A Multi-PDK Dataset for Machine-Learning-Based Post-Layout Capacitance Extraction}

\author{Hector R. Rodriguez}
\orcid{0009-0005-3794-3816}
\affiliation{%
  \institution{Department of Computer Science \& Technology, BNRist, Tsinghua University}
  \city{Beijing}
  \country{China}
}
\email{lad24@mails.tsinghua.edu.cn}

\author{Jiechen Huang}
\orcid{0000-0002-9748-1829}
\affiliation{%
  \institution{Department of Computer Science \& Technology, BNRist, Tsinghua University}
  \city{Beijing}
  \country{China}
}
\email{hjc22@mails.tsinghua.edu.cn}

\author{Wenjian Yu}
\orcid{0000-0003-4897-7251}
\authornote{This work was supported by National Natural Science Foundation of China (Grant No. 62090025). W. Yu is the corresponding author.}
\affiliation{%
  \institution{Department of Computer Science \& Technology, BNRist, Tsinghua University}
  \city{Beijing}
  \country{China}
}
\email{yu-wj@tsinghua.edu.cn}
\begin{abstract}
    We present CapBench, a fully reproducible, multi-PDK dataset for capacitance extraction. The dataset is derived from open-source designs, including single-core CPUs, Systems-on-Chip, and media accelerators. All designs are fully placed and routed using 14 independent OpenROAD flow runs spanning three technology nodes: ASAP7, NanGate45, and Sky130HD. From these layouts, we extract 61,855 3D windows across three size tiers to enable transfer learning and scalability studies. High-fidelity capacitance labels are generated using RWCap, a state-of-the-art random-walk solver, and validated against the industry-standard Raphael, achieving a mean absolute error of 0.64\% for total capacitance. Each window is pre-processed into density maps, graph representations, and point clouds. We evaluate 10 machine learning architectures that illustrate dataset usage and serve as baselines, including convolutional neural networks (CNNs), point cloud transformers, and graph neural networks (GNNs). CNNs demonstrate the lowest errors (1.75\%), while GNNs are up to 41.4$\times$ faster but exhibit the larger errors (10.2\%), illustrating a clear accuracy–speed trade-off. Code and dataset are available at \url{https://github.com/THU-numbda/CapBench}.
\end{abstract}


\begin{teaserfigure}
  \includegraphics[width=1\textwidth]{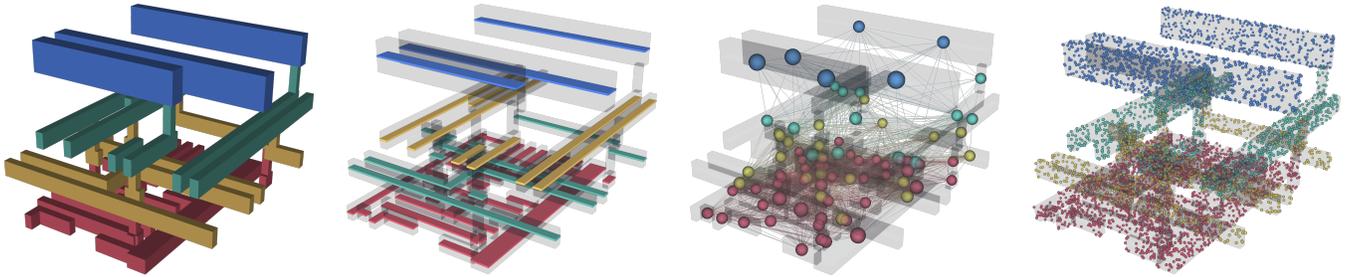}
  \caption{Formats of CapBench windows: 3D blocks, density maps, graphs, and point clouds.}
  \Description{A four-panel teaser showing the same layout window as a 3D conductor view, per-layer density maps, a conductor graph, and a point cloud.}
  \label{fig:teaser}
\end{teaserfigure}
\maketitle
\section{Introduction}

Machine learning offers a promising solution to many computationally intensive tasks in electronic design automation (EDA), such as routing congestion prediction~\cite{zhao2019machine}, chip thermal simulation~\cite{ranade2022thermal}, and parasitic extraction~\cite{yang2022cnn,liu2023gnn,PCT-Cap}. However, the advancement of ML-based techniques is often limited by the availability of high-quality, diverse datasets. An open-source dataset can facilitate research across various EDA tasks, including logic synthesis, placement, routing, and parasitic extraction. In this work, we focus on capacitance extraction, a critical step in post-layout verification that significantly impacts timing and power analysis.

Accurate 3D capacitance estimation is essential for post-layout timing closure and power optimization. Traditional field-solver-based extraction offers high accuracy but is prohibitively slow for large-scale layouts, motivating data-driven surrogates that can predict interconnect capacitances efficiently. These approaches can significantly accelerate the extraction process while maintaining high accuracy~\cite{yang2022cnn,liu2023gnn,PCT-Cap}. However, the lack of publicly available datasets with high-fidelity capacitance labels and diverse layout geometries hinders the effective development and benchmarking of such models.

Despite the growth of open-source hardware~\cite{chipyard,opentitan2020} and open-source EDA tools such as OpenROAD~\cite{ajayi2019openroad}, most process design kits (PDKs) used to generate physical layouts remain proprietary. Given these constraints, prior work often continues to rely on proprietary PDKs, which restricts reproducibility. Even with fully open-source toolchains and open PDKs such as SkyWater130~\cite{skywater130}, NanGate45~\cite{freepdk45}, and ASAP7~\cite{clark2016asap7}, the entry barrier for data generation remains high due to flow complexity and the computational cost of data preparation and labeling.

This need has motivated the creation of open datasets such as CircuitNet~\cite{10158384,jiang2024circuitnet}, ChiPBench~\cite{wang2024benchmarkingendtoendperformanceaibased}, OpenABC-D\cite{chowdhury2021openabc}, ForgeEDA\cite{shi2025forgeeda}, and EDA-schema\cite{shrestha2024eda}. However, they either lack complete post-layout geometric information, or do not include high-fidelity capacitance labels.

To address these challenges, we present CapBench, a multi-PDK dataset for capacitance extraction derived from open-source designs fully placed and routed using the OpenROAD flow across three technology nodes: ASAP7, NanGate45, and Sky130HD. Our contributions are (1) a curated dataset of physical layouts from open-source designs using the three most popular open-source PDKs; (2) generated multi-scale layout windows and employed RWCap~\cite{yu2013rwcap,huang2025efficient,rwcap_v5} and Raphael to generate high-fidelity capacitance labels; and (3) prepared a baseline of three representative ML architectures for 3D capacitance extraction, providing a foundation for future ML-for-EDA research. Fig.~\ref{fig:teaser} shows a sample CapBench window in 3D-layout, density-map, graph, and point-cloud formats.

\section{Preliminaries}

We summarize capacitance extraction, related EDA datasets, and the main ML architectures for 3D capacitance extraction.

\subsection{Capacitance Extraction}

Capacitance extraction aims to determine the electrostatic coupling among conductors in a layout by solving for the potential distribution in the surrounding dielectric. This is commonly formulated as a boundary value problem of Laplace's equation:

\begin{equation}
\begin{cases}\label{eq:laplace}
    \nabla \cdot (\varepsilon \nabla\phi) = 0, & \text{in } \Omega, \\
    \phi = \phi_0, & \text{on } \partial\Omega_D,\\
    \frac{\partial \phi}{\partial \mathbf{n}} = 0, & \text{on } \partial\Omega_N.
\end{cases}
\end{equation}

Given the potential $\phi$, the induced charge on each conductor can be computed, and the capacitance matrix follows from the charge–voltage relationship \( \mathbf{Q} = \mathbf{C}\mathbf{V} \). Numerical 3D field solvers, such as boundary element, finite difference and floating random walk methods, directly approximate this solution and provide high accuracy at significant computational cost.

\subsection{Existing Datasets for ML in EDA}

Several open datasets have been developed to advance ML research in EDA, each addressing different aspects of the design flow such as logic synthesis, placement and routing.

\textit{CircuitNet} \cite{10158384,jiang2024circuitnet} provides over 10,000 samples from open-source RISC-V CPUs and AI chips. It includes grid and graph representations of the layouts with labels for routability, IR-drop, and timing prediction. However, it lacks capacitance data and complete layout geometries, limiting reproducibility and extension to other tasks due to its use of proprietary PDKs.

\textit{ChiPBench} \cite{wang2024benchmarkingendtoendperformanceaibased} focuses on ML-based chip placement evaluation using designs processed through the OpenROAD flow. While it excels in placement evaluation, it does not address parasitic extraction or include capacitance data. \textit{OpenABC-D} \cite{chowdhury2021openabc} uses the OpenROAD toolchain for logic synthesis on open-source IPs. The dataset provides AIG representations and quality-of-result metrics but is limited to pre-layout data and lacks capacitance information.

\textit{ForgeEDA} \cite{shi2025forgeeda} spans logic synthesis through placement, combining 1,189 open-source register-transfer level (RTL) designs ranging from CPUs and arithmetic units to controllers and interfaces. The flow uses both commercial tools (Synopsys Design Compiler and Cadence Innovus) and open-source tools (OpenROAD), all implemented with the Sky130 PDK. The dataset includes RTL, and-inverter graph (AIG), and placed-netlist representations. Moreover, the authors perform a detailed comparison between open-source and commercial tools and demonstrate competitive scalability and fully reproducible flows. However, the dataset is limited to a single PDK and stops at placement, without the routed geometries required for capacitance extraction.

\textit{EDA-schema} \cite{shrestha2024eda} focuses on graph-based representations across the full physical design flow. It uses 20 designs from the IWLS'05 benchmark suite~\cite{albrecht2005iwls}, including controllers, processors, cryptographic, and signal-processing circuits. All designs were synthesized, placed, and routed using OpenROAD with the Sky130HD PDK. The dataset provides unified graph representations across different abstraction levels and includes post-routing parasitic information such as resistances and capacitances extracted via OpenROAD's built-in extractor, OpenRCX~\cite{OpenRCX2021}. While this pattern-matching-based extractor enables fast and consistent labeling, it lacks the precision of field-solver-based capacitance extraction.

\begin{table}[t]
\setlength{\belowcaptionskip}{3pt}
\centering
\caption{Comparison of EDA Datasets for ML}
\vspace{-1em}
\label{tab:dataset_comparison}
\begin{tabular}{@{}l l l l@{}}
\toprule
\textbf{Dataset} & \textbf{PDK} & \textbf{EDA Stage} & \textbf{Outputs} \\
\midrule
CircuitNet & Proprietary &  Complete & Task-Specific Labels \\
ChiPBench & NanGate45 & Complete & Complete Data \\
OpenABC-D & — & Synthesis & AIG \\
ForgeEDA & Sky130HD & Placement & AIG, Placed Netlist \\
EDA-schema & Sky130HD & Complete & Graph, Capacitance \\
\midrule
& Sky130HD, &  & Complete Data, \\
\textbf{CapBench}&  Nangate45, & Complete & High-acc. \\
  &  ASAP7 &  & Capacitance  \\
\bottomrule
\end{tabular}
\end{table}

\subsection{ML Methods for Capacitance Extraction}

Recent advances in machine learning have led to three main architectural approaches for 3D capacitance extraction: convolutional neural networks (CNNs) that treat layouts as image-like density maps, point cloud transformers (PCTs) that sample geometric features from conductor boundaries, and graph neural networks (GNNs) that model interconnect structures as spatial networks. Each approach represents layout geometry differently, leading to distinct trade-offs between computational efficiency and accuracy.

\textit{CNN-Cap \cite{yang2022cnn}} uses a ResNet architecture \cite{he2016resnet} to predict capacitance in both 2D and 3D layouts, representing the conductors as channels with binary masks corresponding to different layers. The authors propose a dual-model approach, with total and coupling capacitances computed by separate models. The 3D dataset was constructed from a single static random-access memory (SRAM) design using 5~$\mu$m$\times$5~$\mu$m windows, creating 8,685 samples. However, the high degree of window overlap (7$\times$ coverage) raises questions about potential data leakage between training and test sets, which may impact generalization estimates. Additionally, the dataset’s focus on only three metal layers and the geometric regularity of SRAM layouts may not capture the full complexity of diverse designs and advanced-node routing patterns.

\textit{PCT-Cap \cite{PCT-Cap}} uses a PCT architecture for capacitance extraction. The capacitance extraction is performed at local level, with 0.5~$\mu$m windows for a 28 nm process technology. The PCT-Cap-10 model, with 10 self-attention layers, achieves an average error of 1.40 on the total capacitance test dataset, compared to ResNet-101's 2.29 error. However, the inference time for PCT-Cap-10 with the same hardware is 1.79$\times$ slower than ResNet-101. Additionally, the three routing layer scope and the relatively small window size may not capture the full complexity of multi-layer coupling effects prevalent in modern designs with extensive inter-layer interactions over larger distances.

\textit{GNN-Cap \cite{liu2023gnn}} uses a spatial-based graph convolution network (GCN) with two message-passing layers to extract the total and coupling capacitance of a circuit window. After the message passing layers, the embeddings are used to extract the total and coupling capacitance using a fully connected linear layer. The network was trained with 400 windows of side length 20~$\mu$m, and then generalized to test cases with up to 274,135 nets, achieving better accuracy than the pattern-matching-based commercial solver StarRC while being 11.43 $\times$ faster. The models showed effective transfer to unseen test cases, though the evaluation methodology's use of training data during testing presents a limitation for assessing true generalization.

\section{Methodology}

\subsection{Toolchain and Data Generation Overview}
\begin{figure}[t]
  \includegraphics[width=0.9\columnwidth]{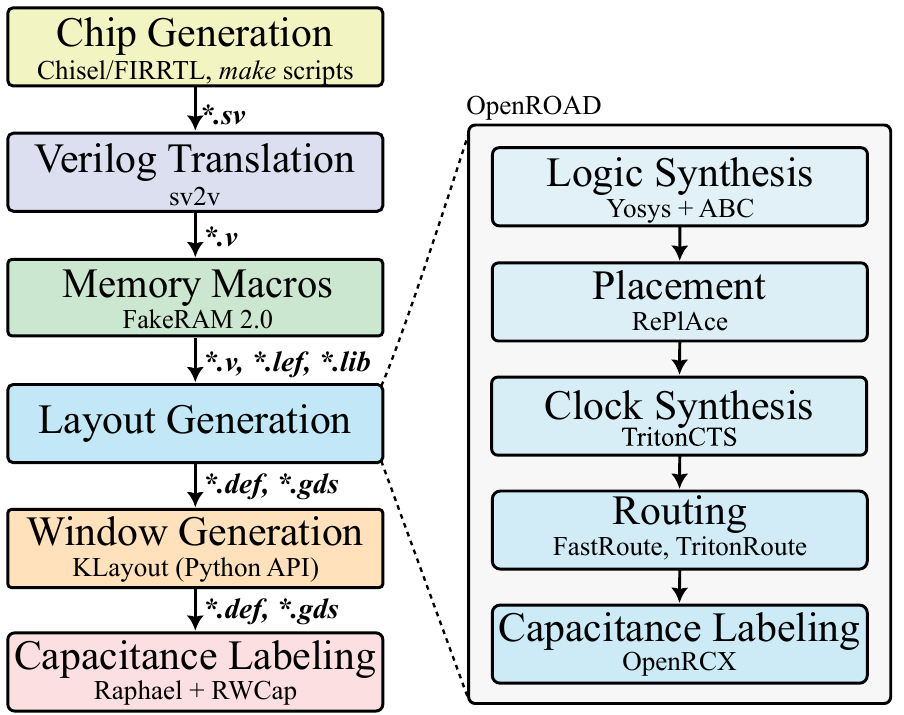}
  \vspace{-0.5em}
  \caption{CapBench dataset generation flow and toolchain.}
  \Description{Pipeline diagram from RTL sources through OpenROAD implementation, KLayout window extraction, and capacitance labeling with RWCap and Raphael validation.}
  \label{fig:flow}
\end{figure}

Figure \ref{fig:flow} summarizes the full pipeline from RTL to labeled layout windows. Designs are generated, translated to Verilog, and augmented with FakeRAM macros. The resulting netlist is passed to OpenROAD, which handles synthesis, placement, CTS, routing, and finally capacitance extraction via OpenRCX. Finished layouts are then cut into spatial windows using KLayout. Each window is labeled using RWCap~\cite{yu2013rwcap}, and the commercial solver Raphael is  used to ensure high-fidelity ground-truth capacitance values.

\subsection{Design and Process Node Selection}

CapBench includes a curated set of designs chosen to reflect realistic and diverse layout structures. Ibex provides a small RV32 core with simple control logic and minimal pipeline depth \cite{ibex_riscv_core}. JPEG introduces a highly regular, deeply pipelined datapath typical of media encoders. CVA6 scales up to a Linux-capable RV64 core with larger caches and wider execution units, stressing routing and clocking \cite{cva6_github}. We also incorporate two Chipyard-based configurations \cite{chipyard}: TinyRocket, a lightweight Rocket core tuned for embedded-class use, and RocketChip-Benchmark, a full SoC instance with TileLink interconnect and multi-level cache hierarchy. Together, these designs span control-centric CPUs, compute pipelines, and full SoC fabrics, ensuring the dataset covers a wide range of physical and architectural characteristics.

We synthesize the designs using the three open-source PDKs. Sky130HD represents a mature 130 nm planar technology with wide metal tracks. NanGate45 models a 45 nm CMOS node with tighter routing and modern design rules. ASAP7 pushes further to a 7 nm FinFET process, introducing multi-patterning, very narrow metal lines, and high-contrast dielectric stacks that strongly affect coupling behavior. Using all three lets us capture variation across older, mainstream, and advanced nodes, which is essential for testing cross-process generalization.

\subsection{Layout and Window Generation}

All designs are implemented with the OpenROAD flow, including synthesis, placement, routing, and parasitic extraction. We configure the flow for realistic but not aggressively tuned layouts: core utilization ranges from 10 up to 60\%, and target clock frequencies span 1 MHz to 833 MHz depending on node and design size. Memory macros are substituted prior to synthesis using process-compatible FakeRAM instances to prevent the inference of flip-flop–based memory structures. Without this substitution, the synthesis tool would map large on-chip memories to thousands of discrete registers, inflating cell count, area, and routing congestion. After routing, OpenRCX produces full-chip parasitics, and the resulting DEF and GDS are passed to a KLayout pipeline for window extraction.

Window extraction uses KLayout's hierarchical region operations, which lets us generate the windows without flattening the full chip. Before accepting a window, we check that the thinnest cross-section of every net within it satisfies the corresponding design rules, which avoids spurious slivers at the window edges. Conductors that do not meet these criteria are removed. We also preserve logical context: net names are inherited from the DEF file whenever possible, detailed polygon shapes are taken from the GDS, and shapes that do not appear in DEF (such as internal metal in standard cells) are kept but assigned generic local names. Connectivity inside each window is reconstructed using KLayout's layout-to-netlist feature; if a routed net is split into multiple disconnected fragments within the window, we assign numeric suffixes so that related segments retain a common base name.

\begin{table}[t]
  \centering
  \caption{Window size and count.}
  \vspace{-1em}
  \begin{tabular}{@{}llrrrr@{}}
  \toprule
  \textbf{PDK} & \textbf{Size} & \textbf{L} ($\mu$m) & \textbf{\# Windows} & \textbf{From} & \textbf{To}\phantom{aa}  \\
  \midrule
  \multirow{3}{*}{ASAP7}
      & Small   & 0.75    & 2,947   & M1 & M7 \\
      & Medium  & 2.5     & 1,846   & M1 & M9 \\
      & Large   & 5.0     &  1,559  & M1 & M9 \\
  \midrule
  \multirow{3}{*}{Nangate45}
      & Small   & 2.0     & 8,109   & metal1 & metal4 \\
      & Medium  & 5.0     & 8,106   & metal1 & metal7 \\
      & Large   & 10.0    & 2,028   & metal1 & metal7 \\
  \midrule
  \multirow{3}{*}{Sky130HD}
      & Small   & 4.5     & 17,668  & li1 & met4 \\
      & Medium  & 10.0    & 18,220  & li1 & met5 \\
      & Large   & 20.0    & 1,372   & li1 & met5 \\
  \bottomrule
  \vspace{0.5em}
  \end{tabular}
    \label{tab:window_characteristics}
\end{table}

We also restrict the metal layer range within each window to avoid extreme vertical separation. In other words, we prevent configurations where conductors on widely separated upper layers are much farther apart vertically than horizontally. This keeps each window balanced in all three dimensions and avoids poorly conditioned capacitance problems dominated by vertical spacing rather than lateral geometry.

\begin{figure}[b]
  \includegraphics[width=0.80\columnwidth]{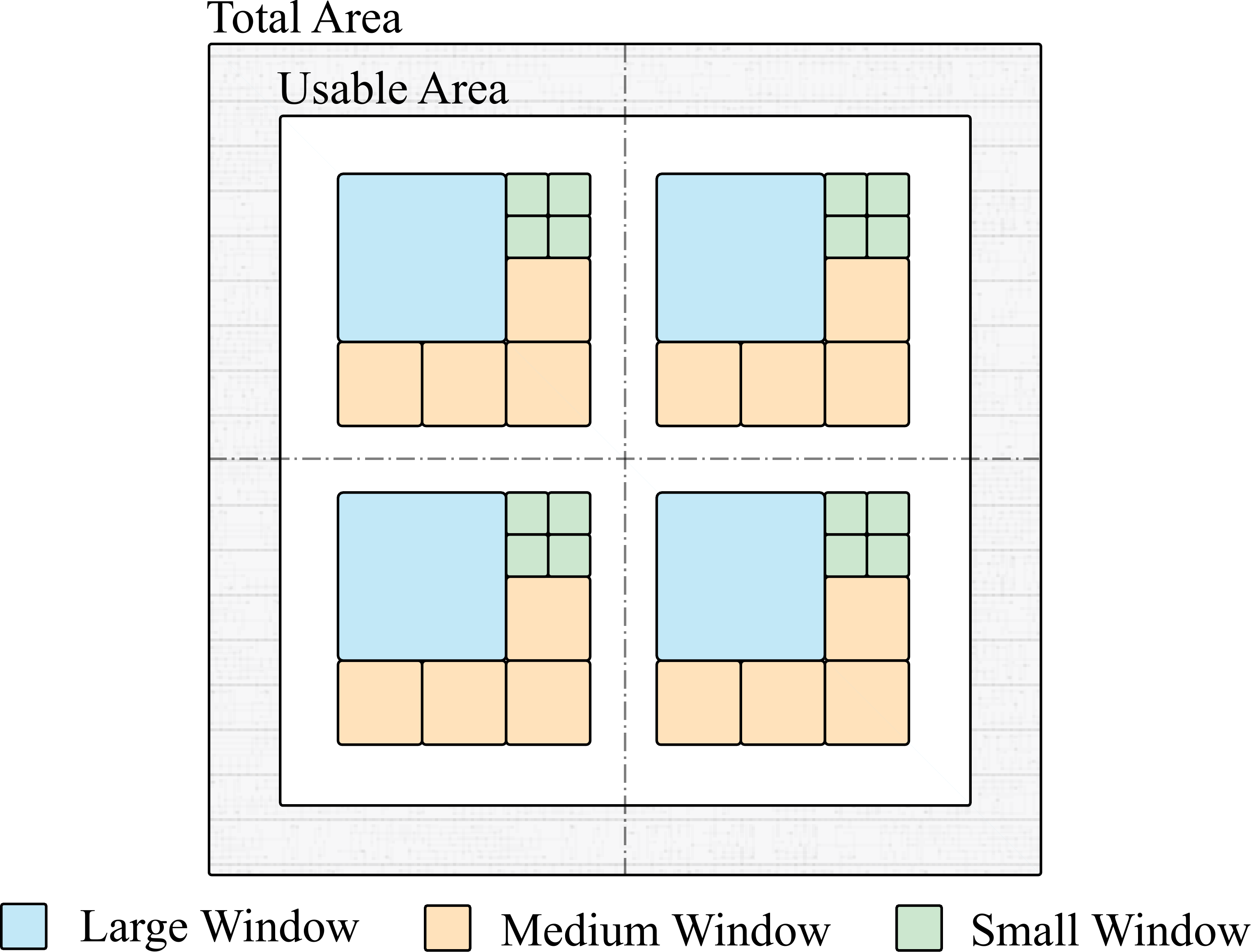}
  \vspace{-0.5em}
  \caption{Window placement on a 2$\times$2 grid.}
  \Description{A 2 by 2 tiling pattern where one large window, four medium windows, and four small windows fill the available area without overlap.}
  \label{fig:window_placement}
\end{figure}

Window placement is illustrated in Fig.~\ref{fig:window_placement}. For each design we first shrink the usable area with a core margin of 10\%, then tile it into a fixed grid of square regions. On average, the selected tiles cover 30.67\% of the core area across designs. Within selected tiles we place a structured set of one large, four medium, and four small windows that touch but do not overlap, following the pattern shown in the figure. Window side lengths are chosen per PDK so that each larger class is at least twice the size of the previous one, which provides multi-scale context while keeping counts manageable. This grid-based placement is designed to avoid spatial leakage between samples, so that train and test sets do not share overlapping layout regions even when they come from the same chip. Total window count and characteristics are listed in Table~\ref{tab:window_characteristics}.

\subsection{Window Capacitance Extraction}

Capacitance extraction was run on a server equipped with two AMD EPYC 7V12 64-Core processors and 512 GB of RAM. RWCap is used as the primary labeling engine due to its scalability and accuracy. As a floating random-walk field solver, RWCap directly estimates 3D capacitances without relying on pattern-matching heuristics or technology-specific rule decks, and it scales more efficiently than finite-difference or finite-element solvers when applied to larger or denser window geometries. Each RWCap job is parallelized across 32 threads, and execution terminates once the variance of the total capacitance estimate falls below 0.5\%, providing tight accuracy bounds with reasonable runtimes across all PDKs and window sizes.

\begin{table}[b]
  \centering
  \caption{RWCap runtime per window.}
  \vspace{-1.0em}
  \label{tab:rwcap_performance}
  \begin{tabular}{@{}llrrr@{}}
  \toprule
  \multirow{2}{*}{\textbf{PDK}} &
  \multirow{2}{*}{\textbf{Size}} &
  \multicolumn{3}{c}{\textbf{Average window}} \\
  \cmidrule(lr){3-5}
  & & \textbf{\# Nets} & \textbf{\# Blocks} & \textbf{Runtime} (s) \\
  \midrule
  \multirow{3}{*}{ASAP7}
      & Small    & 35.5   & 181.4   & 12.35  \\
      & Medium   & 194.3  & 1,873.8 & 98.11  \\
      & Large    & 480.9  & 6,163.6 & 211.10 \\
  \midrule
  \multirow{3}{*}{Nangate45}
      & Small    & 15.0   & 49.8    & 7.10   \\
      & Medium   & 48.9   & 277.3   & 27.48  \\
      & Large    & 135.1  & 1,056.5 & 35.13  \\
  \midrule
  \multirow{3}{*}{Sky130HD}
      & Small    & 15.1    & 82.3   & 25.54 \\
      & Medium   & 39.2    & 380.6  & 39.89 \\
      & Large    & 79.39   & 1,292.8 & 86.90 \\
  \bottomrule
  \end{tabular}
  \vspace{0.5em}
\end{table}

To verify the quality of RWCap-generated labels, we evaluate its results against Raphael across a total of 2,295 small windows. Raphael is the gold standard commercial finite-element field solver, but its higher computational cost makes it unsuitable for large-scale capacitance extraction. Using its default grid, a regrid factor of 2, and a 1\% tolerance target, Raphael required at least two orders of magnitude more runtime than RWCap per window. RWCap achieves mean total-capacitance errors below 1\% and coupling-capacitance errors of roughly 2\% across all three PDKs (Table \ref{tab:raphael_runtime}), providing a level of agreement sufficient for both ML-based modeling and benchmark analysis.

\begin{figure}[t]
    \centering
    \vspace{-0.5em}
    \includegraphics[width=1\columnwidth]{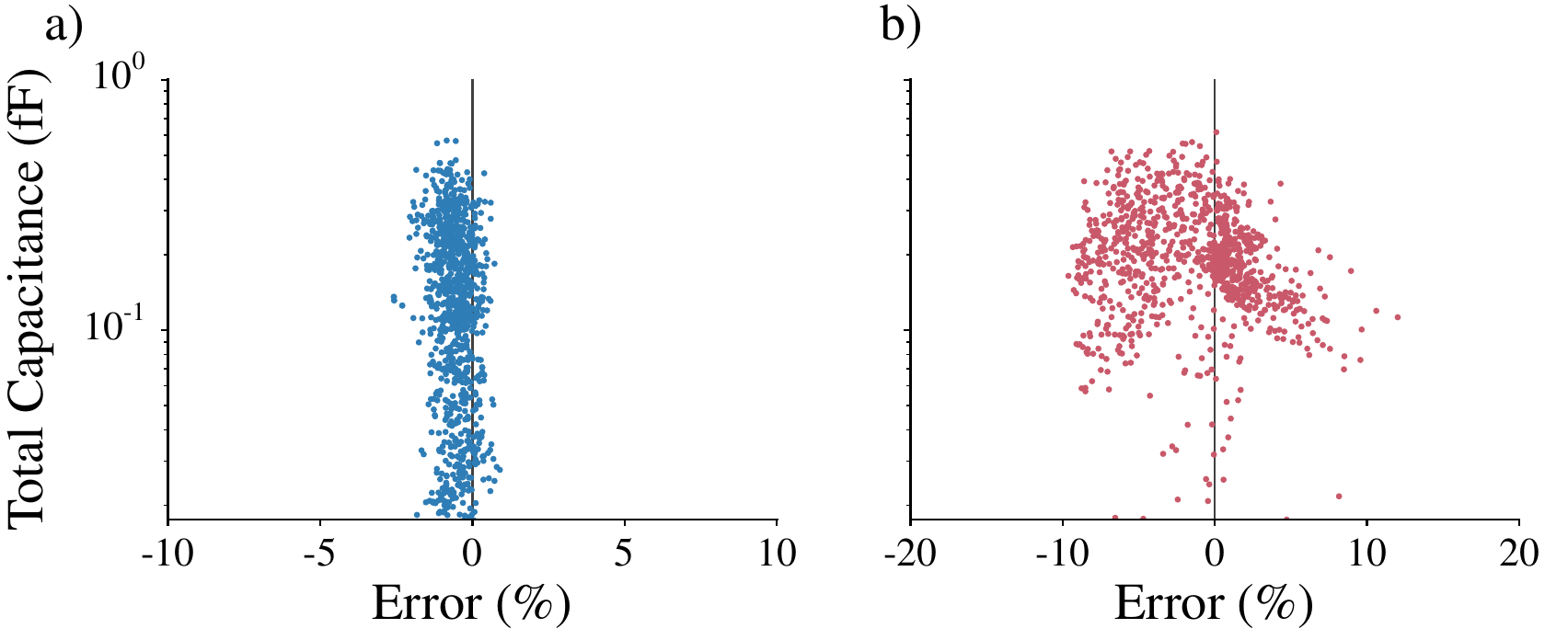}
    \vspace{-2em}
    \caption{RWCap and OpenRCX errors with Raphael as the reference. (a) RWCap, (b) OpenRCX.}
    \Description{Two histograms comparing relative total-capacitance error against Raphael. RWCap errors are tightly concentrated near zero, while OpenRCX errors are broader with visible outliers beyond 20 percent.}
    \label{fig:rwcap_vs_opercx}
\end{figure}

OpenRCX was also evaluated using the default rule decks distributed with OpenROAD. While computationally efficient, its pattern-based approach resulted in substantially higher error rates: on the small-window validation set, it showed mean relative total-capacitance errors close to 20\% for both Sky130HD and Nangate45 nets (see Fig.~\ref{fig:rwcap_vs_opercx}). Although additional calibration may reduce this error, the default configuration is not reliable enough for ground-truth labeling.

\begin{table}[t]
  \centering
  \setlength{\tabcolsep}{2.75pt}
  \caption{RWCap accuracy analysis.}
  \vspace{-1em}
  \label{tab:raphael_runtime}
  \begin{tabular}{@{}lrrrrr@{}}
  \toprule
  \multirow{2}{*}{\textbf{PDK}} & \multirow{2}{*}{\textbf{Samples}} & \textbf{Raphael} & \multicolumn{3}{c}{\textbf{RWCap}}   \\
  \cmidrule(lr){4-6}
  & & ({h/sample})  & Speedup & {Err$_{\text{tot}}$} (\%) & {Err$_{\text{coup}}$}(\%) \\
  \midrule
  ASAP7 & 468 & 3.56 h& 1,038$\times$ & 0.53\% & 1.80\% \\
  Nangate45 & 1,404 & 0.37 h& 189$\times$ & 0.64\% & 1.85\% \\
  Sky130HD & 423 & 10.12 h& 1,426$\times$ & 0.41\% & 1.71\% \\
  \bottomrule
  \end{tabular}
\end{table}

\begin{figure*}[b]
    \centering
    \includegraphics[width=0.9\linewidth]{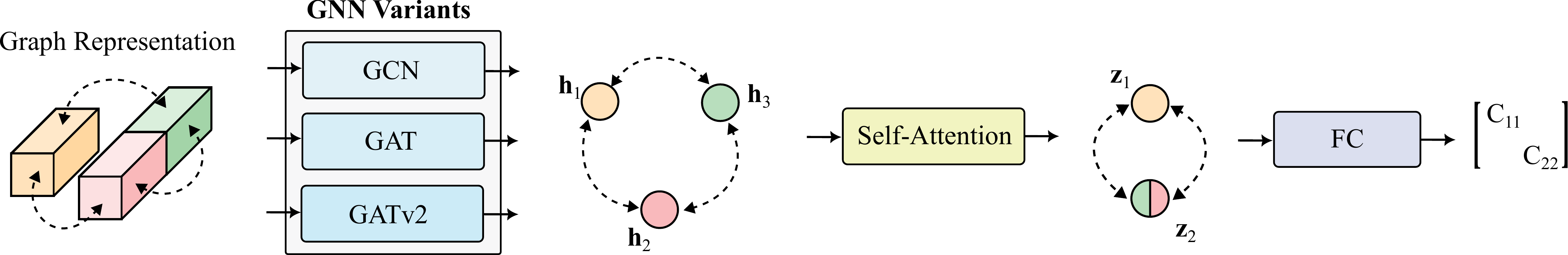}
    \caption{GNN total capacitance prediction flow for a window with two conductors.}
    \Description{A workflow diagram showing conductor decomposition into cuboids, graph construction, message passing, attention-based pooling, and total-capacitance prediction for each net.}
    \vspace{-0.75em}
    \label{fig:gnn_flow}
\end{figure*}

\subsection{Machine Learning Model Baselines}

The CNN-Cap \cite{yang2021cnn, yang2022cnn} architecture employs a dual-model approach for capacitance prediction, with separate models for total capacitance and coupling capacitance. The model utilizes 2D layer density maps as input, where each pixel's value is proportional to the percentage of occupation of its corresponding area by conductive material. In this representation, the master conductor (the conductor for which capacitance is being calculated) is identified by increasing its density values by +1, allowing the network to distinguish it from other conductors. When computing coupling capacitance, CNN-Cap indicates the master conductor following the same approach, and the target conductor is selected by flipping its corresponding density map sign.

Each window is rasterized into a 224×224 grid to match the standard ResNet input resolution. We generate one channel per metal layer and one per via layer, interleaving them to preserve cross-layer spatial relationships. Because different PDKs expose different combinations of metal and via layers, the resulting input tensors have PDK-dependent channel counts. For every configuration, we train ResNet-34, ResNet-50, and ResNet-101.

PCT-Cap formulates the capacitance extraction problem as a point cloud processing task, where the input is a set of 1024 points, randomly sampled from the surfaces of all conductors within the window. Each sampled point is characterized by a vector that encodes its geometric and physical properties, including: (1) the 3D spatial coordinates $(x, y, z)$; (2) the normal vector of the dielectric boundary $(nx, ny, nz)$; (3) the dielectric channel ($\epsilon r$), representing the relative permittivity of the material at that point; and (4) the sign of the electric flux ($\Phi$), an integer channel selection that indicates the potential of the conductor the point belongs to (positive for the master, and negative for the target in coupling calculations). The node features are processed by a PCT architecture, which learns the complex dependencies between all points in the cloud. Following the PCT-Cap formulation, we evaluate models with 4, 6, 8, and 10 self-attention layers to assess the impact of architectural depth on accuracy and throughput.

The GNN model represents interconnect structures as graphs to capture complex spatial relationships for capacitance prediction. The graph construction process begins with the decomposition of conductors into small cuboids. Let $l_m$ denote the maximum allowed length of these cuboids along the x and y directions. Each cuboid becomes a node in the graph, characterized by its geometric features. Standard edges are created between nodes if their spatially extended cuboids overlap, provided their center-to-center distance is less than a threshold $d_e$. To capture long-distance coupling effects that are not represented by standard edges, virtual edges are introduced. These connect nodes in the same or different layers that lack intermediate conductors and have a planar distance less than a threshold $d_{vir}$. Table~\ref{tab:graph_params} lists the parameters used for graph construction for each PDK.

\begin{table}[h]
\centering
\caption{Graph construction.}
\vspace{-1em}
\label{tab:graph_params}
\begin{tabular}{lrrr}
\toprule
\textbf{PDK} & $l_m$ ($\mu$m) & $d_e$ ($\mu$m) & $d_{vir}$ ($\mu$m) \\
\midrule
ASAP7 & 0.5 & 0.25 & 1.0 \\
Nangate45 & 3.0 & 1.5 & 6.0 \\
Sky130HD & 8.0 & 4.0 & 16.0 \\
\bottomrule
\end{tabular}
\end{table}

We implement a GNN that operates on the same underlying training data as other capacitance prediction models, using per-net capacitance labels rather than per-node labels. This architectural choice necessitates an additional aggregation step to transform node-level embeddings into net-level representations, which is a key difference compared to the GNN-Cap implementation. The self-attention aggregation mechanism learns adaptive importance weights for an arbitrary number of nodes within each net. Given node embeddings $\mathbf{h}_i \in \mathbb{R}^d$ for each node $i$ belonging to net $k$, where $d$ is the embedding dimension and $S_k$ denotes the set of all nodes in net $k$, we compute adaptive attention weights and aggregate to the net-level representation:

\begin{equation}
\alpha_i = \mathrm{softmax}_i\left(\mathbf{w}_2^T \, \mathrm{ReLU}\!\left(\mathbf{W}_1 \mathbf{h}_i + \mathbf{b}_1\right) + b_2\right),
\label{eq:attn}
\end{equation}
This attention mechanism computes a learned importance weight $\alpha_i$ for each node and aggregates them into a single net-level representation $\mathbf{z}_k$, which is subsequently used to predict the total capacitance for that net.
\begin{equation}
\mathbf{z}_k = \sum_{i \in S_k} \alpha_i\,\mathbf{h}_i.
\label{eq:pool}
\end{equation}
These two steps form a learned self-attention pooling mechanism that assigns adaptive importance weights to an arbitrary number of nodes within each net and produces the final embedding used to predict total capacitance. This process is illustrated in Fig.~\ref{fig:gnn_flow}.

To enhance the expressivity of the GNN beyond the GCN operator employed in GNN-Cap, we explore attention-based message passing using Graph Attention Networks (GAT) \cite{velickovic2018gat} and the more flexible GATv2 variant \cite{brody2022gatv2}. Both architectures utilize the aggregation mechanism in Eqs.~(\ref{eq:attn}) and (\ref{eq:pool}), but differ in how they compute attention scores. GAT applies a fixed ordering of linear transformations when computing attention logits, which limits its ability to model symmetric interactions between node pairs. GATv2 removes this restriction by allowing the attention mechanism to depend jointly and more flexibly on both source and target node features, resulting in a strictly more expressive and permutation-invariant attention function. This enables GATv2 to better capture geometry-dependent relationships within each net. For consistency with our baseline setup, and to isolate the effect of attention, we do not include the virtual edges used by GNN-Cap to model long-range coupling.

\section{Results}

\subsection{Experimental Setup}

All training and inference experiments were conducted on a server with four NVIDIA RTX 4090 GPUs. For each model, we tested a range of batch sizes and report the maximum stable inference throughput achieved on this hardware.

Our evaluation focuses on the Nangate45 and Sky130HD process nodes to assess model behavior within their intended use cases. All experiments use small windows, which match the scale assumed in the original CNN-Cap and PCT-Cap architectures.

The task for all models is to predict the total capacitance of every conductor within a window. Performance is reported using Mean Absolute Percentage Error (MAPE). We apply an 80/20 train–test split at the window level to avoid geometry leakage, ensuring that all samples from a given window remain in the same split.

All models are trained from scratch for each process node because the architectures learn the technology stack implicitly, including layer counts, thicknesses, and dielectric properties, which vary across nodes. These differences prevent effective weight transfer. For CNN-based models, the number of metal layers also fixes the input channel count, making cross-node reuse infeasible.

To ensure consistency across architectures, all models were trained for 50 epochs, with CNN and PCT models using a batch size of 64. We use the Adam optimizer with a fixed learning rate of 1e-4 and train using Mean Relative Squared Error (MRSE), while reporting Mean Absolute Percentage Error (MAPE) as the primary evaluation metric. Let
$y_i$ denote the ground-truth capacitance for sample $i$, $\hat{y}_i$ its predicted capacitance, and $N$ the number of samples in the corresponding split. MRSE and MAPE are defined as:

\begin{equation}
\mathrm{MRSE} = \frac{1}{N}\sum_i \left(\frac{\hat{y}_i - y_i}{y_i}\right)^2, \qquad
\mathrm{MAPE} = \frac{1}{N}\sum_i \left|\frac{\hat{y}_i - y_i}{y_i}\right|
\end{equation}

\subsection{Results of ML Models}

\begin{table}[t]
  \centering
  \setlength{\tabcolsep}{2pt}
  \setlength{\abovecaptionskip}{5pt}
  \setlength{\belowcaptionskip}{0pt}
  \caption{Performance comparison of ML architectures on small windows.
  MAPE is computed for total capacitance.}
  \label{tab:small_model_performance}
  \begin{threeparttable}
  \begin{tabular}{llrrrr}
  \toprule
  \multirow{2}{*}{\textbf{Model}} & \multirow{2}{*}{\textbf{Arch}} & \multirow{2}{*}{\textbf{Params}} &
  \textbf{Speed\tnote{1}}\phantom{a} & \textbf{Sky130HD} & \textbf{Nangate45} \\
  & & &  {sample/s}  & {Tot. Err} (\%) & {Tot. Err} (\%) \\
  \midrule
  CNN & ResNet-34  & 21.3 M& 4602 & 1.99 & 4.93 \\
  CNN & ResNet-50  & 23.5 M & 2520 & 1.75 & 3.86 \\
  CNN & ResNet-101 & 42.5 M & 1654 & 1.86 & 4.01 \\
  \midrule
  PCT & 4 Layers  & 1.85 M& 4334  & 7.98  & 9.12  \\
  PCT & 6 Layers   & 2.25 M& 3014 & 6.78  & 7.48  \\
  PCT & 8 Layers  & 2.64 M& 2285  & 7.78  & 7.41  \\
  PCT & 10 Layers & 3.03 M& 1862  & 8.75  & 7.81  \\
  \midrule
  GNN & GCN   & 20.9 K & 1470 & 13.9 & 21.4 \\
  GNN & GAT   & 84.4 K & 1741 & 12.1 & 18.0 \\
  GNN & GATv2 & 167.3 K & 1314 & 10.2 & 14.7 \\
  \bottomrule
  \end{tabular}
  \begin{tablenotes}
  \small
  \item[1] Since GNNs predict for all conductors per window while CNNs/PCTs predict per conductor, the GCN is effectively 41.4$\times$ faster than ResNet-34 for an average window with 15.1 conductors.
  \end{tablenotes}
  \end{threeparttable}
\end{table}

The results in Table \ref{tab:small_model_performance} show that CNN-based models achieve the lowest total-capacitance error across both Sky130HD and Nangate45. ResNet-50 offers the best overall accuracy, reaching 1.75\% and 3.86\% on the two nodes. These models benefit from structured 2D density representations that capture spatial conductor patterns with high fidelity. Their accuracy, however, comes at the cost of increased parameter counts and reduced throughput as depth increases, as seen with ResNet-101. In practice, these larger models remain practical on modern GPUs.

PCTs achieve lower memory usage and significantly smaller model sizes, but their error remains noticeably higher. A likely cause is that the standard PCT-Cap setting of 1024 points per window does not provide enough geometric resolution. As a result, the additional expressive power of deeper PCT models is not fully leveraged, and accuracy does not consistently improve with depth. Moreover, although PCT models are more memory efficient, their inference speed is lower than that of the CNN models due to the self-attention layers, which introduce computational costs not present in convolutional backbones.

GNNs show the highest total-capacitance error in Table \ref{tab:small_model_performance}. This is partly expected, since predicting a full-net capacitance value provides a weaker training signal compared to node-level supervision used in prior GNN-Cap approaches. Even so, the progression from GCN to GAT and GATv2 highlights the benefit of increased expressivity and attention mechanisms, with GATv2 reducing error by several percentage points while remaining compact. In terms of efficiency, GNNs achieve the highest effective throughput because they process all conductors in a window simultaneously, which makes them suitable for large-scale inference despite their lower accuracy.

\section{Conclusion}
We introduced CapBench, a fully reproducible multi-PDK dataset for 3D capacitance extraction derived from complete OpenROAD layouts across process technologies Sky130HD, Nangate45, and ASAP7. Along with high-fidelity RWCap labels and multiple window representations, CapBench enables a consistent evaluation of CNN, PCT, and GNN architectures. Our experiments establish baseline accuracy and throughput levels, and they help clarify the trade-offs between accuracy, model size, and inference efficiency under a standardized setup. By releasing the dataset, code, and evaluation scripts, CapBench provides a foundation on which future work can build improved architectures and cross-node generalization methods.

\bibliographystyle{ACM-Reference-Format}
\bibliography{references/reference}

@misc{ajayi2019openroad,
  title={OpenROAD: Toward a Self-Driving, Open-Source Digital Layout Implementation Tool Chain},
  author={Ajayi, T and Blaauw, D and Chan, TB and Cheng, CK and Chhabria, VA and Choo, DK and Coltella, M and Dobre, S and Dreslinski, R and Foga{\c{c}}a, M and others},
  year={2019},
  howpublished={Government Microcircuit Applications \& Critical Technology Conference (GOMACTech)},
  note={Conference paper, pp. 1105--1110},
  url={https://par.nsf.gov/biblio/10171024-openroad-toward-self-driving-open-source-digital-layout-implementation-tool-chain}
}

@article{clark2016asap7,
  title={ASAP7: A 7-nm finFET predictive process design kit},
  author={Clark, Lawrence T and Vashishtha, Vinay and Shifren, Lucian and Gujja, Aditya and Sinha, Saurabh and Cline, Brian and Ramamurthy, Chandarasekaran and Yeric, Greg},
  journal={Microelectronics Journal},
  volume={53},
  pages={105--115},
  year={2016},
  publisher={Elsevier}
}

@inproceedings{yang2021cnn,
  title={CNN-Cap: Effective convolutional neural network based capacitance models for full-chip parasitic extraction},
  author={Yang, Dingcheng and Yu, Wenjian and Guo, Yuanbo and Liang, Wenjie},
  booktitle={2021 IEEE/ACM International Conference On Computer Aided Design (ICCAD)},
  pages={1--9},
  year={2021},
  organization={IEEE},
  publisher={IEEE},
  address={Piscataway, NJ, USA}
}

@article{yang2022cnn,
  title={CNN-Cap: Effective Convolutional Neural Network Based Capacitance Models for Interconnect Capacitance Extraction},
  author={Yang, Dingcheng and Li, Haoyuan and Yu, Wenjian and Guo, Yuanbo and Liang, Wenjie},
  journal={ACM Transactions on Design Automation of Electronic Systems (TODAES)},
  year={2023},
  volume={28},
  number={4},
  articleno={56},
  numpages={22},
  month={may},
  doi={10.1145/3564931},
  publisher={Association for Computing Machinery}
}

@INPROCEEDINGS{PCT-Cap,
  author={Cai, Ye and Liang, Yuyao and Luo, Zhipeng and Xie, Biwei and Li, Xingquan},
  booktitle={2024 2nd International Symposium of Electronics Design Automation (ISEDA)},
  title={PCT-Cap: Point Cloud Transformer for Accurate 3D Capacitance Extraction},
  year={2024},
  pages={421-426},
  publisher={IEEE},
  address={Piscataway, NJ, USA},
  doi={10.1109/ISEDA62518.2024.10617673}
}

@article{liu2023gnn,
  title={GNN-Cap: Chip-Scale interconnect capacitance extraction using graph neural network},
  author={Liu, Lihao and Yang, Fan and Shang, Li and Zeng, Xuan},
  journal={IEEE Transactions on Computer-Aided Design of Integrated Circuits and Systems},
  volume={43},
  number={4},
  pages={1206--1217},
  year={2023},
  publisher={IEEE}
}

@ARTICLE{10158384,
  author={Chai, Zhuomin and Zhao, Yuxiang and Liu, Wei and Lin, Yibo and Wang, Runsheng and Huang, Ru},
  journal={IEEE Transactions on Computer-Aided Design of Integrated Circuits and Systems},
  title={CircuitNet: An Open-Source Dataset for Machine Learning in VLSI CAD Applications with Improved Domain-Specific Evaluation Metric and Learning Strategies},
  year={2023},
  volume={42},
  number={12},
  pages={5034--5047},
  publisher={IEEE},
  doi={10.1109/TCAD.2023.3287970}
}

@misc{jiang2024circuitnet,
  title={Circuitnet 2.0: An advanced dataset for promoting machine learning innovations in realistic chip design environment},
  author={Jiang, Xun and Zhao, Yuxiang and Lin, Yibo and Wang, Runsheng and Huang, Ru and others},
  year={2024},
  howpublished={International Conference on Learning Representations (ICLR 2024)}
}

@article{huang2025efficient,
  title={Efficient FRW Transitions via Stochastic Finite Differences for Handling Non-Stratified Dielectrics},
  author={Huang, Jiechen and Yu, Wenjian},
  journal={IEEE Transactions on Computer-Aided Design of Integrated Circuits and Systems},
  year={2026},
  volume={45},
  number={4},
  pages={1746--1750},
  doi={10.1109/TCAD.2025.3602737},
  publisher={IEEE}
}

@article{yu2013rwcap,
  title={RWCap: A floating random walk solver for 3-D capacitance extraction of very-large-scale integration interconnects},
  author={Yu, Wenjian and Zhuang, Hao and Zhang, Chao and Hu, Gang and Liu, Zhi},
  journal={IEEE Transactions on Computer-Aided Design of Integrated Circuits and Systems},
  volume={32},
  number={3},
  pages={353--366},
  year={2013},
  publisher={IEEE}
}

@inproceedings{zhao2019machine,
  title={Machine learning based routing congestion prediction in FPGA high-level synthesis},
  author={Zhao, Jieru and Liang, Tingyuan and Sinha, Sharad and Zhang, Wei},
  booktitle={2019 Design, Automation \& Test in Europe Conference \& Exhibition (DATE)},
  pages={1130--1135},
  year={2019},
  doi={10.23919/DATE.2019.8714724},
  publisher={European Design and Automation Association},
  address={Leuven, Belgium}
}

@inproceedings{ranade2022thermal,
  title={A thermal machine learning solver for chip simulation},
  author={Ranade, Rishikesh and He, Haiyang and Pathak, Jay and Chang, Norman and Kumar, Akhilesh and Wen, Jimin},
  booktitle={Proceedings of the 2022 ACM/IEEE Workshop on Machine Learning for CAD (MLCAD '22)},
  pages={111--117},
  year={2022},
  doi={10.1145/3551901.3556484},
  publisher={Association for Computing Machinery},
  address={New York, NY, USA}
}

@misc{skywater130,
  title = {SkyWater 130 nm Open PDK},
  author = {{SkyWater Technology Foundry} and Google},
  year = {2020},
  howpublished = {\url{https://github.com/google/skywater-pdk}},
  note = {Accessed: 2025-05-01}
}

@misc{freepdk45,
  title = {FreePDK45: An Open-Source Predictive Process Design Kit},
  author = {{North Carolina State University EDA Lab}},
  year = {2008},
  howpublished = {\url{https://www.eda.ncsu.edu/wiki/FreePDK45:Contents}},
  note = {Accessed: 2025-05-01}
}

@misc{chowdhury2021openabc,
  title={Openabc-d: A large-scale dataset for machine learning guided integrated circuit synthesis},
  author={Chowdhury, Animesh Basak and Tan, Benjamin and Karri, Ramesh and Garg, Siddharth},
  year={2021},
  eprint={2110.11292},
  archivePrefix={arXiv},
  primaryClass={cs.LG},
  howpublished={CoRR abs/2110.11292},
  url={https://arxiv.org/abs/2110.11292}
}

@misc{shi2025forgeeda,
  title={ForgeEDA: A Comprehensive Multimodal Dataset for Advancing EDA},
  author={Shi, Zhengyuan and Li, Zeju and Ma, Chengyu and Zhou, Yunhao and Zheng, Ziyang and Liu, Jiawei and Pan, Hongyang and Zhou, Lingfeng and Li, Kezhi and Zhu, Jiaying and others},
  year={2025},
  eprint={2505.02016},
  archivePrefix={arXiv},
  primaryClass={cs.AR},
  howpublished={CoRR abs/2505.02016},
  url={https://arxiv.org/abs/2505.02016}
}

@inproceedings{shrestha2024eda,
  title={EDA-schema: A graph datamodel schema and open dataset for digital design automation},
  author={Shrestha, Pratik and Aversa, Alec and Phatharodom, Saran and Savidis, Ioannis},
  booktitle={Proceedings of the Great Lakes Symposium on VLSI 2024 (GLSVLSI '24)},
  pages={69--77},
  year={2024},
  doi={10.1145/3649476.3658718},
  publisher={Association for Computing Machinery},
  address={New York, NY, USA}
}

@misc{albrecht2005iwls,
  title={IWLS 2005 benchmarks},
  author={Albrecht, Christoph},
  year={2005},
  howpublished={International Workshop on Logic and Synthesis (IWLS)}
}

@misc{wang2024benchmarkingendtoendperformanceaibased,
      title={Benchmarking End-To-End Performance of AI-Based Chip Placement Algorithms},
      author={Zhihai Wang and Zijie Geng and Zhaojie Tu and Jie Wang and Yuxi Qian and Zhexuan Xu and Ziyan Liu and Siyuan Xu and Zhentao Tang and Shixiong Kai and Mingxuan Yuan and Jianye Hao and Bin Li and Yongdong Zhang and Feng Wu},
      year={2024},
      eprint={2407.15026},
      archivePrefix={arXiv},
      primaryClass={cs.AR},
      url={https://arxiv.org/abs/2407.15026},
}

@misc{OpenRCX2021,
  author       = {{The OpenROAD Project}},
  title        = {OpenRCX: A Parasitics Extraction Tool Within OpenDB},
  year         = {2021},
  howpublished = {\url{https://github.com/The-OpenROAD-Project/OpenRCX}},
  note         = {GitHub repository}
}

@misc{velickovic2018gat,
  title     = {Graph Attention Networks},
  author    = {Veličković, Petar and Cucurull, Guillem and Casanova, Arantxa and Romero, Adriana and Lio, Pietro and Bengio, Yoshua},
  year      = {2018},
  howpublished = {International Conference on Learning Representations (ICLR 2018)},
  url       = {https://arxiv.org/abs/1710.10903}
}

@misc{brody2022gatv2,
  title     = {How Attentive are Graph Attention Networks?},
  author    = {Brody, Shaked and Alon, Uri and Yahav, Eran},
  year      = {2022},
  howpublished = {International Conference on Learning Representations (ICLR 2022)},
  url       = {https://arxiv.org/abs/2105.14491}
}

@inproceedings{he2016resnet,
  title     = {Deep Residual Learning for Image Recognition},
  author    = {He, Kaiming and Zhang, Xiangyu and Ren, Shaoqing and Sun, Jian},
  booktitle = {Proceedings of the IEEE Conference on Computer Vision and Pattern Recognition (CVPR)},
  year      = {2016},
  pages     = {770--778},
  doi       = {10.1109/CVPR.2016.90},
  publisher = {IEEE Computer Society},
  address   = {Los Alamitos, CA, USA}
}

@misc{cva6_github,
  author       = {{OpenHW Group and PULP Platform}},
  title        = {{CVA6: An Application-Class 64-bit RISC-V Core}},
  year         = {2024},
  howpublished = {\url{https://github.com/openhwgroup/cva6}},
  note = {Accessed: 2025-11-19}
}

@article{chipyard,
  author={Amid, Alon and Biancolin, David and Gonzalez, Abraham and Grubb, Daniel and Karandikar, Sagar and Liew, Harrison and Magyar,   Albert and Mao, Howard and Ou, Albert and Pemberton, Nathan and Rigge, Paul and Schmidt, Colin and Wright, John and Zhao, Jerry and Shao, Yakun Sophia and Asanovi\'{c}, Krste and Nikoli\'{c}, Borivoje},
  journal={IEEE Micro},
  title={Chipyard: Integrated Design, Simulation, and Implementation Framework for Custom SoCs},
  year={2020},
  volume={40},
  number={4},
  pages={10-21},
  doi={10.1109/MM.2020.2996616},
  ISSN={1937-4143},
}

@misc{ibex_riscv_core,
  author       = {{lowRISC Contributors}},
  title        = {{Ibex: 32-bit In-order RISC-V Core Implemented in SystemVerilog}},
  year         = {2025},
  howpublished = {\url{https://github.com/lowRISC/ibex}},
  note         = {Open-source RISC-V core. Accessed: 2026-04-13}
}

@Misc{rwcap_v5,
	author = {Numbda},
  title = {\emph{RWCap5-v5} \url{https://numbda.cs.tsinghua.edu.cn/download.html}},
  year = {2025}
}

@misc{opentitan2020,
  title        = {Open{T}itan: Open {S}ource {S}ilicon {R}oot of {T}rust},
  author       = {OpenTitan Project},
  year         = {2020},
  howpublished = {\url{https://opentitan.org}},
  note         = {Project website}
}

\end{document}